\documentclass[runningheads,a4paper]{llncs}

\usepackage{amssymb}
\usepackage{graphicx}

\usepackage{url}

\newcommand{\un}[1]{\ensuremath{\unskip\,\mathrm{#1}}}
\urldef{\mailsa}\path|igoychuk@uni-potsdam.de |    
\newcommand{\keywords}[1]{\par\addvspace\baselineskip
\noindent\keywordname\enspace\ignorespaces#1}
\newcommand{\chem}[1]{\ensuremath{\mathrm{#1}}}
\begin{document}

\mainmatter  % start of an individual contribution

\title{Stochastic modeling of excitable dynamics: \\
improved Langevin model for \\
mesoscopic channel noise}

\titlerunning{Improved Langevin model for mesoscopic channel noise}

\author{Igor Goychuk}
%\thanks{Supported by the DFG (German Research Foundation), Grant GO 2052/1-2}
%}
%
\authorrunning{Stochastic modeling of excitable dynamics}

\institute{Institute for Physics and Astronomy, University of Potsdam\\
Karl-Liebknecht-Str. 24/25, 14476 Potsdam-Golm, Germany \\
\mailsa\\
\url{http://www.physik.uni-augsburg.de/~igor/}}


\toctitle{Lecture Notes in Computer Science}
\tocauthor{}
\maketitle

\begin{abstract}

Influence of mesoscopic channel noise on excitable dynamics of living cells
became a hot subject within the last decade,
and the traditional biophysical models of neuronal dynamics 
such as Hodgkin-Huxley model have been generalized to incorporate such effects. 
There still exists but  a controversy on how to do it in a proper and
computationally efficient way. Here we introduce an improved Langevin description
of stochastic Hodgkin-Huxley dynamics with natural boundary conditions 
for gating variables. 
It consistently describes the channel noise variance in a good agreement with discrete state
model. 
Moreover, we show by comparison
with our improved Langevin model that two earlier Langevin models by Fox and Lu also work
excellently starting from several hundreds of ion channels 
upon imposing numerically reflecting boundary conditions for gating
variables.

\end{abstract}

\section{Introduction}

Hodgkin-Huxley (HH) model of neuronal excitability \cite{HH52} provides a milestone for
biophysical understanding of information processing in living systems \cite{Koch}
in terms of electrical spikes mediated by
ionic currents through voltage-dependent membrane pores made
by ion channel proteins. One considers the cell membrane as an insulator
with specific electrical capacitance $C_m$ per unit of area, which is perforated by
ionic channels providing generally voltage-dependent parallel ionic pathways 
with specific conductances $G_i$ per unit of area for various sorts of ion channels. This yields 
the following equation for transmembrane electrical potential difference $V$
\begin{equation}
   \label{voltage-eq}
   C_m \frac{dV}{dt} +G_{\chem{K}}(n) (V-E_{\chem{K}})
   +G_{\chem{Na}}(m,h) ( V - E_{\chem{Na}})
    +G_{\chem{L}} (V - E_{L}) = I_{\rm ext}\, .
\end{equation}
 Here, three ionic currents are taken into account,
sodium $\chem{Na}$, potassium $\chem{K}$ and unspecific leakage current (mainly due to 
chloride ions). This is nothing else the Kirchhoff current law, which takes into account the ionic
and capacitance currents, as well as an external current $I_{\rm ext}$ which
can stimulate electrical excitations. This equation reflects assumption on Ohmic conductance of
completely open ion channels with $E_i$ being the
reversal or Nernst potentials. They emerge due to the difference of ionic concentrations inside and 
outside of the excitable cell, which are kept approximately constant by the work of ionic pumps,
which is not considered explicitely.  
%(kept approximately constant in spite of ionic current flows
%by the work of ionic pumps charging continuously the ``batteries''). 
Nonlinearity comes from the  open-shut gating
dynamics of sodium
and potassium channels. The corresponding specific conductances
\begin{eqnarray}
  \label{conductances}
  G_{\chem{K}}(n)&&=g_{\chem{K}}^{\mathrm{max}}n^{4}(V,t) , \nonumber \\
  G_{\chem{Na}}(m,h)&&=g_{\chem{Na}}^{\mathrm{max}}m^{3}(V,t) h(V,t)\, ,
\end{eqnarray}
depend on three voltage-dependent gating variables, $n$, $m$, and $h$, where $n(t)$ is the 
probability 
of one gate of potassium channel to be open (more precisely the fraction of open gates), 
$m$ corresponds to one activation gate of sodium 
channel, and $h$ is the fraction of closed sodium inactivation gates. 
One assumes four independent
identical gates for potassium channel, hence its opening probability is $n^4$, as well as
 three activation and one inactivation gate for the sodium channel. Hence, $m^3h$ is the
 fraction of open sodium channels.
 The maximal conductances $g_{\chem K}^{\mathrm{max}}$ and $g_{\chem{Na}}^{\mathrm{max}}$
 can be expressed via the unitary conductances $g_{i,0}$ of single ion channels as 
 $g_{i}^{\mathrm{max}}=g_{i,0}\rho_i$, where $\rho_i$ is the membrane density of the ion
 channels of sort $i$. The gating dynamics is in turn described by the relaxation kinetics
 \begin{equation}
  \label{gates}
  \frac{d}{dt} x =
  \alpha_{x}(V)\ (1-x)-\beta_{x}(V)\ x , \quad x=m,h,n\, ,
\end{equation}
%(fractions of open and closed gates add to one) 
with voltage-dependent %opening ($\alpha_x$) and closing ($\beta_x$) 
rates
\begin{eqnarray}
  \label{rates1}
  \alpha_{m}(V) &&= \frac{ 0.1\ ( V + 40 )}{1-\exp [ - ( V + 40 ) /
    10] },\;
  \beta_{m}(V) = 4 \ \exp [ - ( V + 65 ) / 18 ]\, ,  \\
  \alpha_{h}( V ) &&=  0.07 \ \exp [ - ( V + 65 ) / 20 ], \;
  \beta_{h}( V ) = \{ 1 + \exp [ - ( V + 35 ) / 10 ] \}^{-1}\, , \\
  \alpha_{n}( V ) &&= \frac{ 0.01 \ ( V + 55 )}{ 1 - \exp [ -( V +
    55 )/10 ]},\;
  \label{eq:rates2}
  \beta_{n}( V ) = 0.125 \ \exp [ - ( V + 65 ) / 80 ]\, .
\end{eqnarray}
Here the voltage is measured in millivolts and rates in inverse milliseconds. Other
classical parameters of HH model suitable to describe excitable dynamics of squid giant
axon are:  $C_m=1 \un{\mu F/cm^2}$, $E_{\chem{Na}}=50\un{mV}$, $E_{\chem{K}}=-77\un{mV}$,
$E_{\mathrm{L}}=-54.4\un{mV}$, $G_{\mathrm{L}} =0.3\un{mS/cm^2}$, 
$g_{\chem K}^{\mathrm{max}}=36\un{mS/cm^2}$, $g_{\chem{Na}}^{\mathrm{max}}=120\un{mS/cm^2}$.
 
The set of four coupled nonlinear differential equations defined by (\ref{voltage-eq})-
(\ref{eq:rates2})
presents a milestone in biophysics and neuroscience because of its very clear and insightful 
physical background. In the same spirit, one can build up various other 
conductance-based biophysical 
models starting from the pertinent molecular background and following to the 
bottom-up approach.
However, it assumes macroscopically large numbers of ion channels in neglecting completely
the mesoscopic channel noise effects. The number of ion channels in any cell is, however, finite,
and the corresponding channel noise can be substantial \cite{Koch}. 
Especially, one confronts with this question by
considering the spatial spike propagation among approximately piece-wise
isopotential membrane clusters of ion channels \cite{Anna}.

\section{Stochastic Hodgkin-Huxley equations}

How to take stochastic dynamics of ion channels within the physical framework of HH model into account
is subject of numerous studies \cite{Koch,Strassberg,FoxLu}. The most rigorous way is to consider 
the variable population 
of open ion channels as a birth-and-death process \cite{Kampen}.  
Consider for simplicity a population of $N$ 
independent two-state ion channels (one gate only) 
with opening rate
$\alpha$ and closing rate $\beta$ (constant under voltage clamp). Each ion channel 
fluctuates dichotomously between the closed state with zero conductance and 
the open state having unitary conductance $g_0$. 
For such two-state Markovian channels, the stationary probability of opening is 
$p_0=\alpha/(\alpha+\beta)$ and the averaged conductance is 
$\langle g(t)\rangle=p_0g_0$. 
The number of 
open channels $n$ is binomially distributed
with probability $P_{N}^{\rm st}(n)=p_0^n(1-p_0)^{N-n}N!/(n!(N-n)!)$, average $\langle n\rangle =p_0N$,
and variance $\langle (n-\langle n\rangle)^2\rangle=Np_0(1-p_0)=N\alpha\beta/(\alpha+\beta)^2$. 
For sufficiently large 
$N\geq 100$, we introduce quasi-continuous variable $0\leq x(t)=n(t)/N\leq 1$, with smoothened 
binomial probability density 
$p_{N}^{\rm st}(x)=Np^{xN}(1-p)^{N(1-x)}N!/(\Gamma(1+xN)\Gamma(1+(1-x)N))$. 
Use of approximate Stirling
formula $n!\approx (n/e)^n$ yields
\begin{eqnarray}\label{stat}
p^{\rm st}_N(x)\approx C_N(\alpha,\beta)\left(\frac{\alpha}{x}\right)^{Nx}
\left(\frac{\beta}{1-x}\right)^{N(1-x)}\;,
\end{eqnarray}
where $C_N(\alpha,\beta)$ is a normalization constant.
We are looking for the best diffusional (continuous) approximation for discrete state 
birth-and-death
process defined by the master equation
\begin{eqnarray}
\dot P_N(n)&&=F(n-1)P_N(n-1)+B(n+1)P_N(n+1)-[F(n)+B(n)]P_N(n)
\end{eqnarray}
for $1\leq n\leq N-1$, with forward rate $F(n)=\alpha (N-n)$ and backward rate $B(n)=\beta n$,
complemented by the boundary conditions
\begin{eqnarray} 
\dot P_N(0)&&=B(1)P_N(1)-F(0)P_N(0), \\
\dot P_N(N)&&=F(N-1)P_N(N-1)-B(N)P_N(N)\;.
\end{eqnarray}

\subsection{Diffusional approximations for birth-and-death process}

\subsubsection{Kramers-Moyal expansion and standard diffusional approximation.}
 A standard way to obtain diffusional approximation for $p_N(x):=P_N(x N)/\Delta x$ 
 ($\Delta x=1/N$) with rates $f(x):=F(xN)\Delta x$, $b(x):=B(xN) \Delta x$ 
 is to do the Kramers-Moyal expansion \cite{Kampen,Hanggi82}, like $p_N(x+\Delta x)\approx
 p_N(x)+(\partial p_N(x)/\partial x)\Delta x+
 (\partial^2 p_N(x)/\partial x^2)(\Delta x)^2/2$, 
  $f(x+\Delta x)\approx
 f(x)+(df(x)/dx)\Delta x+(d^2f(x)/dx^2)(\Delta x)^2/2$,
 to the second order. This yields the Fokker-Planck equation 
 \begin{eqnarray}
 \frac{\partial }{\partial t} p(x,t)=-\frac{\partial  }{\partial x}[f(x)-b(x)]p(x,t)+
 \frac{\partial^2  }{\partial x^2}D_{\rm KM}(x)p(x,t)
 \end{eqnarray}
with diffusion coefficient $D_{\rm KM}(x)=[f(x)+b(x)]/(2N)$.
This Fokker-Planck equation corresponds to the Langevin equation
\begin{eqnarray}
\dot x=f(x)-b(x)+\sqrt{2D_{\rm KM}(x)}\xi(t),
\end{eqnarray}
where $\xi(t)$ is white Gaussian noise of unit intensity, $\langle\xi(t)\xi(t') \rangle=\delta(t-t') $,
in pre-point, or Ito interpretation \cite{Gard}. This equation is quite general for any one-dimensional
 birth-and-death process
within this standard diffusional approximation.
% This is a particular example of the so-called
%chemical Langevin equation by Gillespie.
For the considered population of ion channels,
\begin{eqnarray}\label{standard}
\dot x =\alpha(1-x)-\beta x+\sqrt{[\alpha(1-x)+\beta x]/N}\xi(t).
\end{eqnarray}
This is stochastic equation for a gating variable in the stochastic generalization
of Hodgkin-Huxley equations by Fox and Lu \cite{FoxLu}. It replaces Eq. (\ref{gates}) with corresponding
voltage-dependent $\alpha_x(V)$, $\beta_x(V)$, and $N=N_{\rm Na}=\rho_{\rm Na}S$ for
$m$ and $h$, or $N=N_{\rm K}=\rho_{\rm K}S$ for the variable $n$. $S$ is the area of membrane
patch, and $\rho_{\rm Na}=60\mu m^{-2}$, $\rho_{\rm K}=18\mu m^{-2}$ within HH model \cite{Strassberg}.
Clearly, in the limit $N\to\infty$ the channel noise vanishes, restoring the deterministic HH model.
We name this model the second model by Fox and Lu 
(Fox-Lu 2) in application to stochastic HH dynamics.
\subsubsection{Linear noise approximation.}
The further approximation  (Fox-Lu 1 within stochastic HH model) 
is obtained by 
$D_{\rm KM}(x)\to D_{\rm KM}(x_{\rm eq})=const$, where $x_{\rm eq}$ is equilibrium
point of deterministic dynamics, $f(x_{\rm eq})=b(x_{\rm eq})$. 
It corresponds to the so-called $1/\Omega$ expansion with linear additive noise approximation 
advocated by van Kampen \cite{Kampen}. Then, with $x_{\rm eq}=p_0=\alpha/(\alpha+\beta)$ 
Eq. (\ref{standard}) reduces to
\begin{eqnarray}
\dot x =\alpha(1-x)-\beta x+\sqrt{2\alpha\beta/[N(\alpha+\beta)]}\xi(t).
\end{eqnarray}
\subsubsection{Diffusional approximation with natural boundaries.}
The both diffusional approximations are not quite satisfactory because they do not
guarantee  the boundary conditions in a natural way. As a result,
 for a sufficiently small opening 
probability $p_0\ll 1$, and not sufficiently large number of channels the negative values, $x<0$,
become possible with appreciable probabilities $p(x,t)>0$. Likewise, the larger
than one values, $x>1$, are also possible when the opening probability $p_0$ is close to one.
 However, this deficiency can 
easily be corrected numerically by imposing  reflecting boundary conditions at $x=0$ and $x=1$
in stochastic simulations. With this correction, Langevin approximation of stochastic HH dynamics
is widely used \cite{Schmid01,Schmid04a,Schmid06,Anna}. However, it is not quite clear if this
procedure indeed delivers the correct results \cite{FNL}. To clarify the issue, we consider
a different diffusional approximation with natural reflecting boundaries
which naturally bound stochastic dynamics to the interval $0\leq x \leq1$.

For this,  we first demand that the diffusional approximation is consistent with the stationary
distribution of  birth-and-death process, which  can be expressed 
as $P_N^{\rm st}(n)= \exp [-\Phi(n)]P_N^{\rm st}(0)$ in terms
of a pseudo-potential $\Phi(n)=-\sum_{i=1}^{n}\ln\left [F(i-1)/B(i) \right]$ \cite{Kampen}.
Hence, in the continuous limit, $p_N^{\rm st}(x)\propto \exp [-N\phi(x)]$, with
pseudo-potential $\phi(x)=-\int_0^x \ln\left [f(x')/b(x') \right]d x'=
\ln(1-x)-x\ln(\alpha(1-x)/(x\beta))$. This
indeed yields the probability density (\ref{stat}).
The corresponding Fokker-Planck equation must read
\begin{eqnarray}
 \frac{\partial }{\partial t} p(x,t)=
 \frac{\partial  }{\partial x}\left (D(x)e^{-N\phi(x)}
 \frac{\partial  }{\partial x} e^{N\phi(x)}p(x,t)\right )\\
 =\frac{\partial  }{\partial x}ND(x)\phi'_x(x)p(x,t)+\frac{\partial  }{\partial x}D(x)
 \frac{\partial  }{\partial x}p(x,t)
 \end{eqnarray}
with 
\begin{eqnarray}
ND(x)\phi'_x(x)=b(x)-f(x) \;
 \end{eqnarray}
in order to be also consistent with the deterministic limit $N\to\infty$. The last
equation fixes the diffusion coefficient as
\begin{eqnarray}
D(x)=\frac{1}{N}\frac{f(x)-b(x)}{\ln [f(x)/b(x)]} \;.
\end{eqnarray}
The Langevin equation which corresponds to this best diffusional approximation of the birth-and-death
processes \cite{Hanggi82,Hanggi84}
reads
\begin{eqnarray}
\dot x=f(x)-b(x)+\sqrt{2D(x)}\xi(t),
\end{eqnarray}
in the post-point, or Klimontovich-H\"anggi interpretation \cite{Hanggi82}. 
In the standard 
Ito interpretation suitable for integration with stochastic Euler algorithm \cite{Gard}
 the corresponding Langevin equation becomes
\begin{eqnarray}
\dot x=f(x)-b(x)+D'_x(x)+\sqrt{2D(x)}\xi(t) \;
\end{eqnarray}
with spurious drift $D'_x(x)$.
In application to stochastic dynamics of one gating variable it reads
\begin{eqnarray}
\dot x=\alpha(1-x)-\beta x+D'_x(x)+\sqrt{2D(x)}\xi(t) \;.
\end{eqnarray}
 with
\begin{eqnarray}
D(x)=\frac{1}{N}\frac{\alpha(1-x)-\beta x}{\ln [\alpha(1-x)/(\beta x)]} \;.
\end{eqnarray}
Replacing with such equations the stochastic equations for gating variables
in the standard Langevin variant of stochastic Hodgkin-Huxley equations we obtain
the improved Langevin description of mesoscopic channel noise, with natural boundaries
because  $D(0)=D(1)=0$, i.e. the channel noise (and the probability flux)
vanishes exactly at the reflecting boundaries, in the theory. Nevertheless, in numerical algorithm
one must yet additionally secure such boundaries for any \textit{finite} integration time step $\delta t$.
Notice also that near the equilibrium point with
$|f(x)-b(x)|\ll f(x)+b(x)$, $D(x)\approx D_{\rm KM} (x)$, and the standard diffusional
approximation is almost restored, almost, if to neglect the spurious drift correction $D'_x(x)$, 
which still remains within the Ito interpretation.

\begin{figure}
\centering
\includegraphics[height=4.3cm]{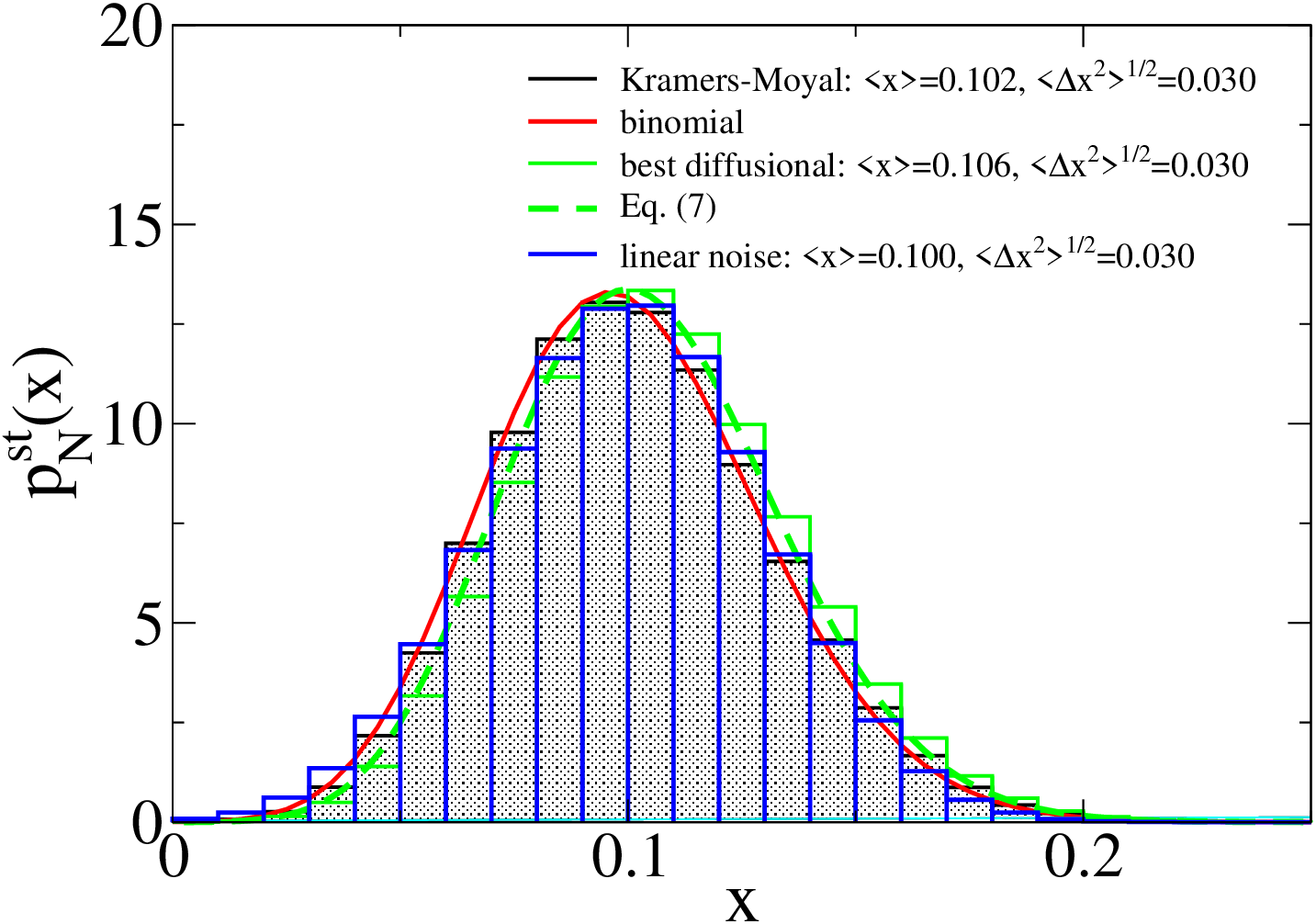}\hfill
\includegraphics[height=4.3cm]{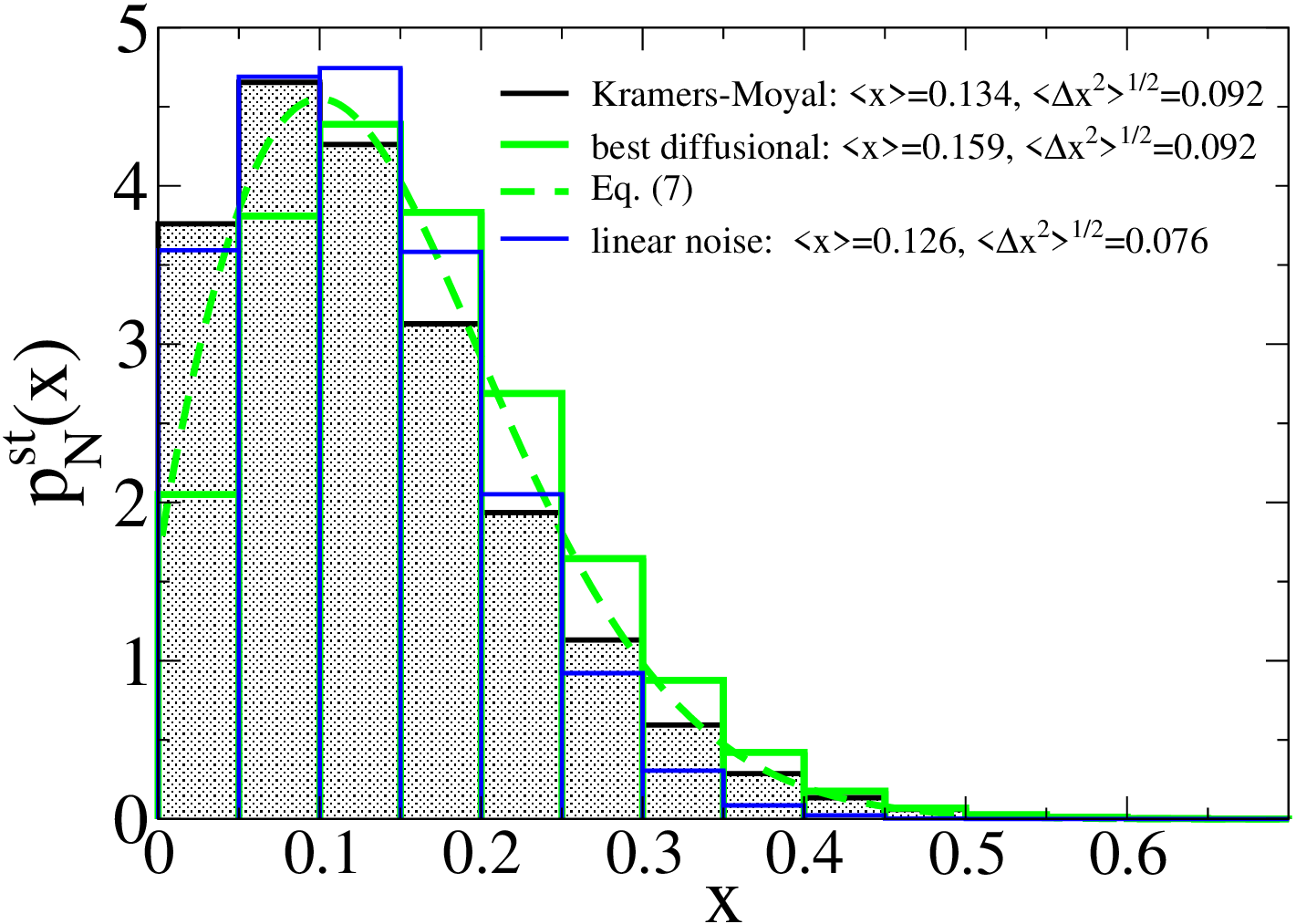}
\caption{Stationary distributions of gating variable $x$ for two ensembles of ion
channels with $\alpha=1$ and $\beta=9$, (a) 
$N=100$ and (b) $N=10$. Numerics are compared with binomial distribution (a) and distribution
 (\ref{stat}) for the best diffusional approximation }
\label{Fig1}
\end{figure}

We test the best diffusional approximation for a gating variable 
against the earlier Langevin descriptions
with reflecting boundary conditions implemented numerically. 
For this we use stochastic Euler algorithm with time step $\delta t=0.001$
for several values of $N$ and the simulation software XPPAUT \cite{Bard}. The results
are shown for $\alpha=1$ and $\beta=9$ with $p_0=0.1$ in Fig. \ref{Fig1} for
 $N=100$ (a) and $N=10$ (b). As a big surprise, the simplest linear noise approximation 
 actually seems to work best, if only to implement reflecting boundary conditions.
 For $N=100$, it  reproduces well the still somewhat skewed binomial distribution with the exact
 mean $\langle x \rangle=0.1$ and standard deviation 
 $\langle \Delta x^2\rangle^{1/2}=0.03$. Even for $N=10$, it gives the mean closer
 to the correct value of $0.1$ within the discrete state model. However, the variance
 then deviates from the theoretical value $\langle \Delta x^2\rangle^{1/2}\approx 
 0.095$ larger than within two other approximations. For a 
 sufficiently large $N=1000$ (not shown), all three diffusional approximations give practically
 identical results, within the statistical errors of simulations.
 Surprisingly, all three work reasonably well even for $N=10$!
 However, such a performance is \textit{a priori} not guaranteed for stochastic nonlinear
 dynamics with voltage-dependent $\alpha(V(t))$ and $\beta(V(t))$. In fact, for a multistable
 dynamics  the best diffusional approximation
 is generically expected \cite{Hanggi84} to operate essentially better.
 
\begin{figure}
\centering
\includegraphics[height=4.2cm]{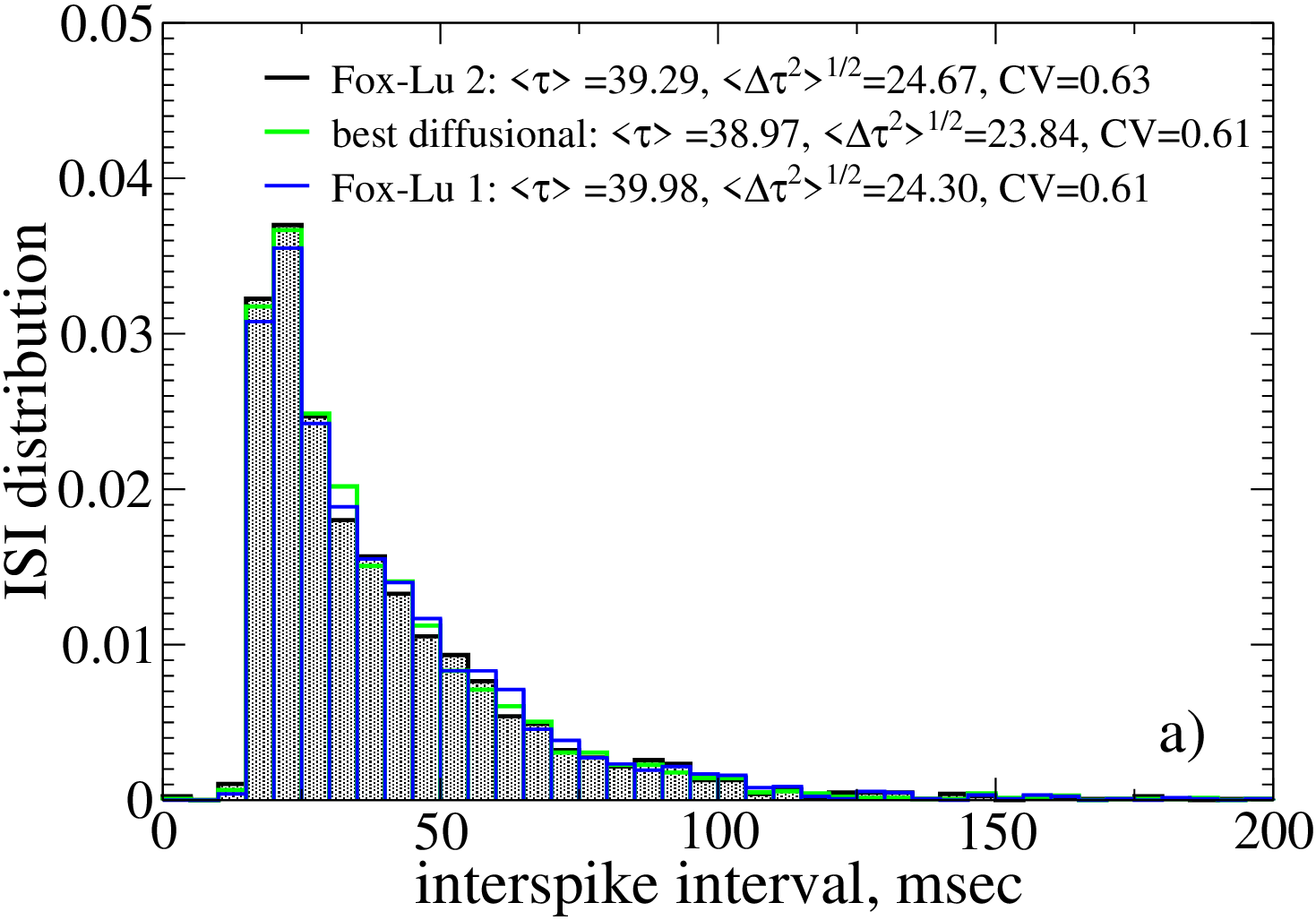}\hfill
\includegraphics[height=4.2cm]{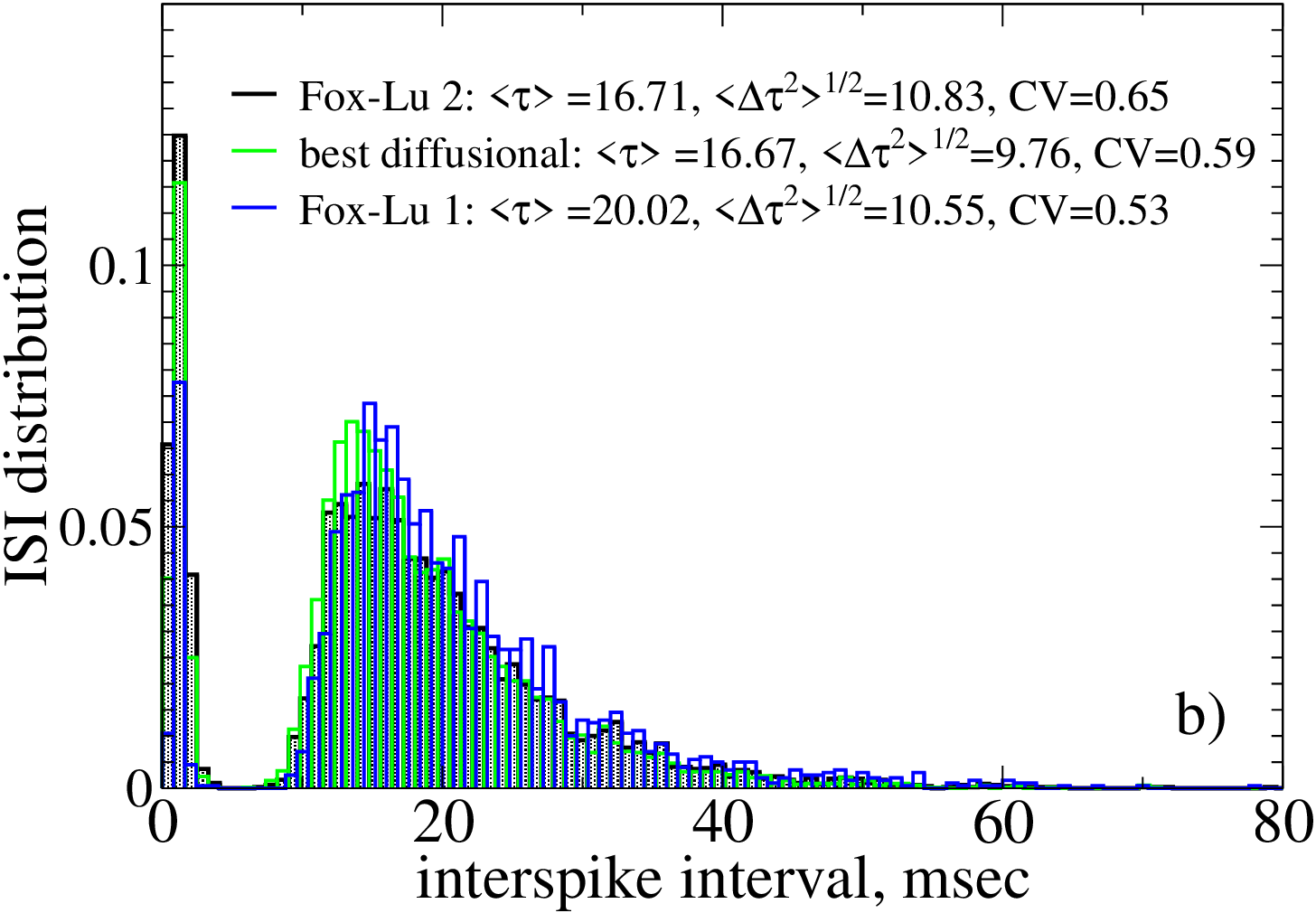}
\caption{ Interspike time interval distribution for self-excitable dynamics, $I_{\rm ext}=0$, due to the
channels noise for two membrane patches: (a) $S=10\mu m^2$ ($N_{\rm Na}=600$,
$N_{\rm K}=180$), and (b) $S=1\mu m^2$ ($N_{\rm Na}=60$,
$N_{\rm K}=18$). }
\label{Fig2}
\end{figure}

We compare three different Langevin descriptions of stochastic HH dynamics in Fig. \ref{Fig2},
for two different membrane patches. Here, the interspike interval distributions
are presented, together with the corresponding mean, $\langle \tau\rangle$,
standard deviation, $\langle [ \tau-\langle \tau\rangle]^2\rangle^{1/2}$, and
the relative standard variation, or the coefficient of variation, 
$C_V=\langle [ \tau-\langle \tau\rangle]^2\rangle^{1/2}/\langle \tau\rangle$,
which measures the spike coherence. For $S=10\;\mu m^2$, all three approximations
agree well. However, for $S=1\; \mu m^2$ the discrepancies become apparent, and
we prefer our improved Langevin description on general theoretical grounds.

\begin{figure}
\vspace{0.5cm}
\centering
\includegraphics[height=4.2cm]{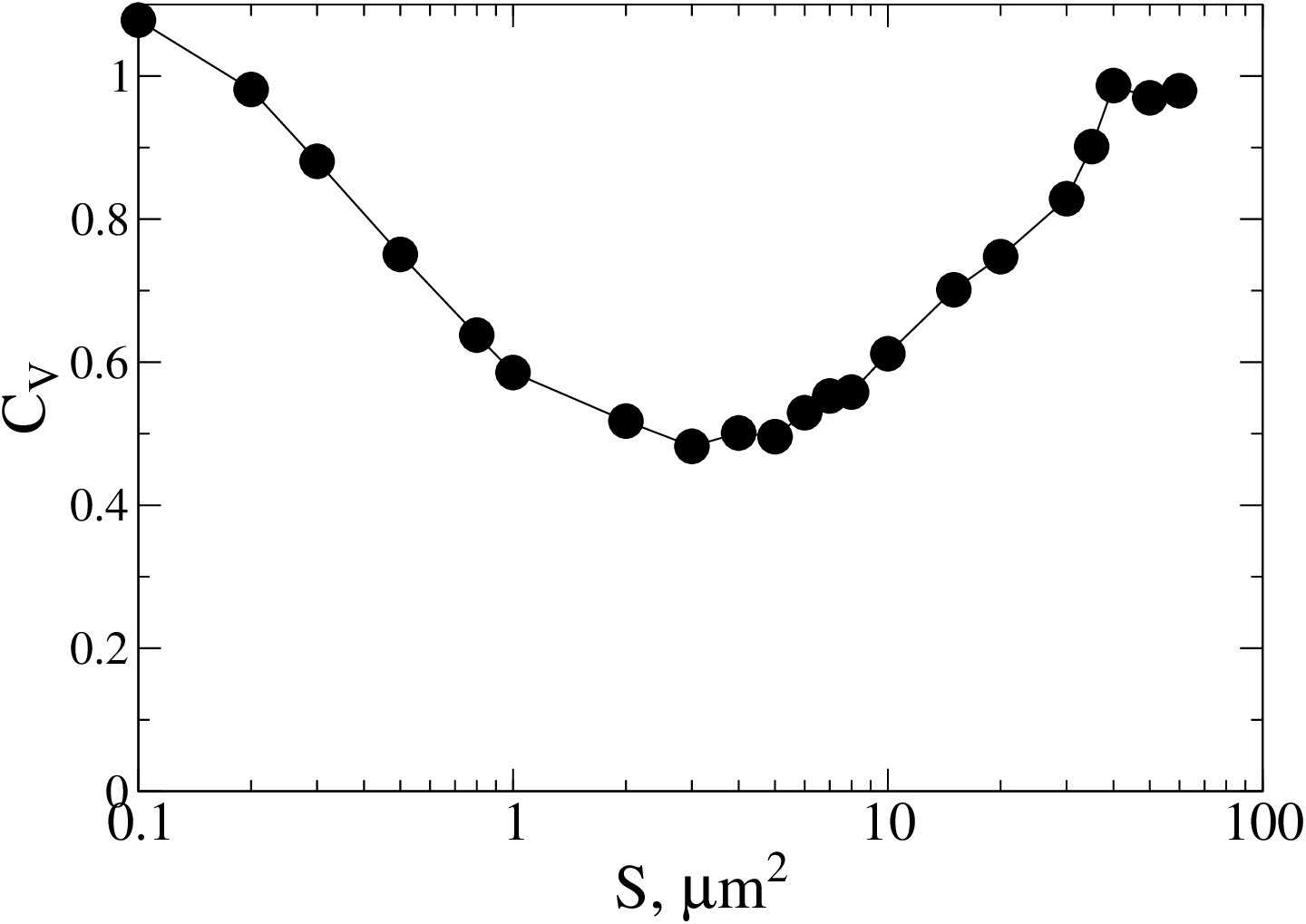}
\caption{ Coefficient of variation versus the membrane patch size within our variant
of stochastic HH model. Self-excitable dynamics, $I_{\rm ext}=0$.}
\label{Fig3}
\end{figure}

The coefficient of variation $C_V$, calculated within
our Langevin variant of stochastic HH model, 
is plotted \textit{vs.} the patch size $S$ in Fig. \ref{Fig3}. It
 displays a typical 
coherence resonance \cite{Pikovsky} behavior revealed earlier within stochastic HH models 
in \cite{Schmid01,Jung} as a system-size coherence resonance. There exists an optimal patch size
(optimal number of ion channels) with most coherent stochastic dynamics due
to internal mesoscopic noise.

\subsection{Summary and Conclusions}

In this paper, we presented the best diffusional Langevin approximation for excitable cell dynamics within
stochastic Hodgkin-Huxley model, with natural boundary conditions for the channel noise implemented.
It has clear theoretical advantages over the standard diffusional approximation in the case
of transitions induced by mesoscopic noise as discussed for bistable birth-and-death processes
long time ago \cite{Hanggi84}. However, within stochastic HH model for a sufficiently large number
of ion channels, the standard diffusional approximations were shown to work also very good.
Hence, this work confirmes the validity of the previous work done within the Langevin approximations
of stochastic HH dynamics,
for a sufficently large number of channels.
This does not mean, however, that for other excitable models  the situation will not be changed.
Generally, the improved Langevin description should operate better.
 Other stochastic models of excitable dynamics, e.g.
stochastic Morris-Lecar model can easily be improved accordingly. This task, as well as comparison
with discrete state stochastic models for channel noise, is left for a future investigation.

\subsubsection*{Acknowledgments.} 

Support by the DFG (German Research Foundation), Grant GO 2052/1-2, is gratefully acknowledged.

%\section{The References Section}\label{references}

\end{document}